\documentclass{sig-alternate-05-2015}

\usepackage[T1]{fontenc}
\usepackage{xspace}
\usepackage{graphicx}
\usepackage{multirow}

\newcommand{\projecttitle}{{\scshape LocationSafe}\xspace}

\makeatletter
\def\@copyrightspace{\relax}
\makeatother

\begin{document}


\title{LocationSafe: Granular Location Privacy for IoT Devices}

%
%
%
%
%

\numberofauthors{3} 
%
\author{
%
%
\alignauthor
Joshua Joy\\
       \email{jjoy@cs.ucla.edu}
\alignauthor
Minh Le\\
       \email{im1nh27@ucla.edu}
\alignauthor
Mario Gerla\\
       \email{gerla@cs.ucla.edu}
}

\maketitle

\begin{abstract}

Today, mobile data owners lack consent and control over the release and utilization of their location data. Third party applications continuously process and access location data without data owners granular control and without knowledge of how location data is being used. The proliferation of IoT devices will lead to larger scale abuses of trust.

In this paper we present the first design and implementation of a privacy module built into the GPSD daemon. The GPSD daemon is a low-level GPS interface that runs on GPS enabled devices. The integration of the privacy module ensures that data owners have granular control over the release of their GPS location. We describe the design of our privacy module and then evaluate the performance of private GPS release and demonstrate that strong privacy guarantees can be built into the GPSD daemon itself with minimal to no overhead.


\end{abstract}
\section{Introduction}

Today data owners' personal mobile devices are constantly being tracked and monitored by third party applications without data owners granular consent and control. Data owners' trust is being continuously violated~\cite{facebooklocation}.

Data owners have a desire to occasionally share their location data, though desire granular control and approved consent. Third party analysts seek to track data owners continuously. Unfortunately today this tension has resulted in disproportionate control being in favor of the third party analysts.  

\begin{figure}[t!]
\includegraphics[width=1\columnwidth]{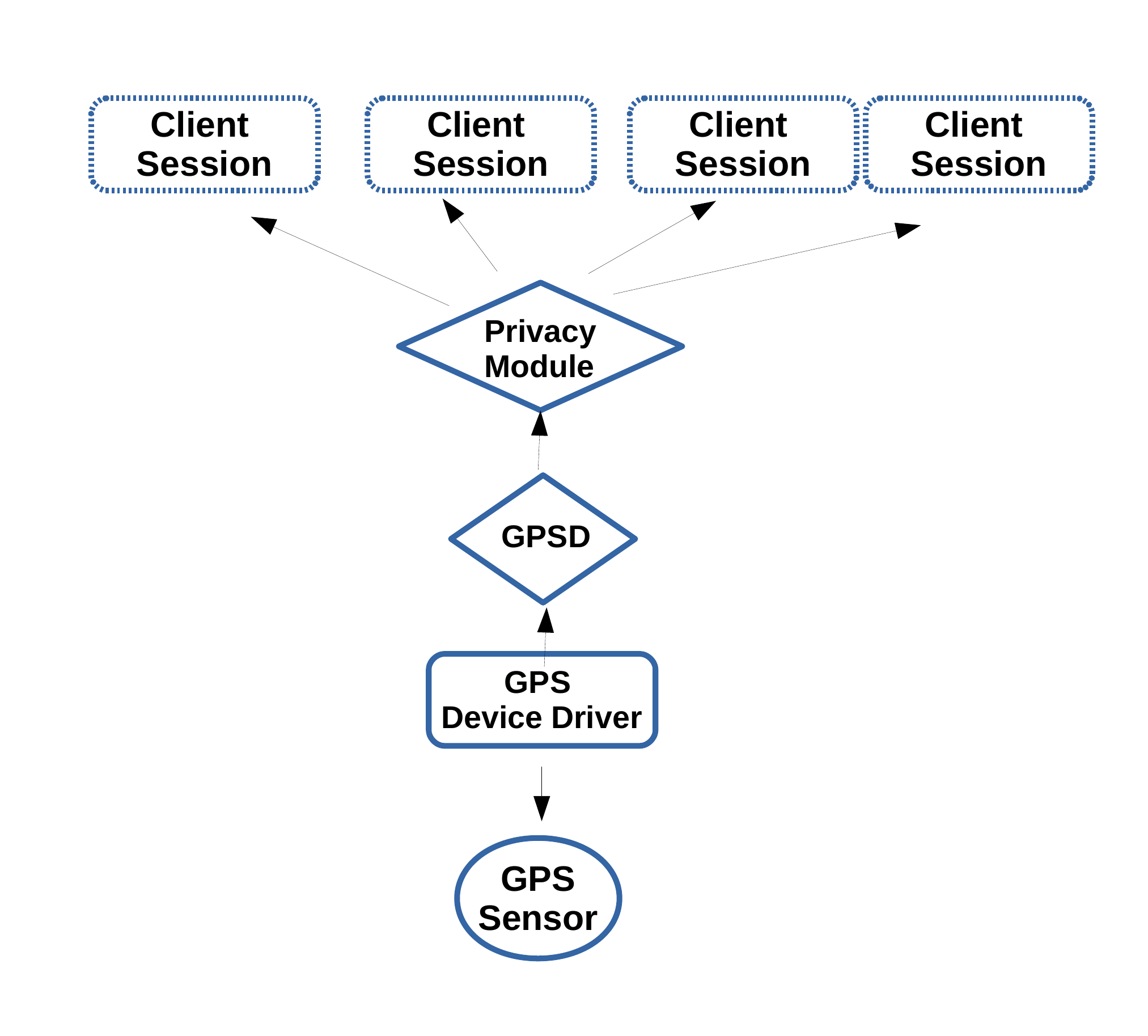}
\caption{Privatization occurs before data is released to the client application.}
\label{fig:gpsd}
\end{figure}

Recent research has tried to improve user behavior in recognizing permission issues \cite{DBLP:conf/soups/FeltHEHCW12}, user-defined runtime constraints \cite{DBLP:conf/ccs/NaumanKZ10}, or tools to help developers identify least-privilege \cite{VCC:W2SP11}.

Additionally, permission managers (e.g., Android and iOS) offer binary permissions to disable or enable location services. However, while this allows data owners to disable location services for applications that do not require location (e.g., Flashlight application) \cite{flashlight}, fine grained granularity is still missing. An Android modification called CynagonMod has a module called XPrivacy~\cite{xprivacy}. XPrivacy enables data owners to configure random or a static location, empty cell ID, blocks geofences from being set, prevents sending NMEA data to application, prevents cell tower updates from being sent to an application, prevents aGPS, returns empty Wi-Fi scans, and disables activity recognition. Ultimately, this provides the data owner control at the application layer. 

User applications requesting data of users is a binary permission, either I share my data or I don't. However, sensitive data such as location needs finer control on how accurate and how often the location information is released. Users should be able to control the granularity of their personal data that is released. Users require freedom and control over their own personal data.

However, these approaches discards several important facts: 1) these privacy mechanisms protect at the application layer only and the underlying operating system still has access to all system location APIs 2) granular privacy permission solutions (e.g., XPrivacy) are only for rooted Android phones 3) there is no compromise between third party analyzers and data owners. The expected proliferation of IoT devices will further exacerbate these privacy issues. 

In this paper, we present the first (to our knowledge) implementation of a privacy module to GPSD. Figure~\ref{fig:gpsd} shows an overview of the flow of queries and responses and demonstrates that the privatization occurs before releasing the data back to the application. The privacy module ensures that all GPS data is released according to the data owner's consent and choice. We demonstrate that appropriate methodologies can be placed which provides strong location privacy guarantees, yet enable analyzers access to privatized location data.

\begin{enumerate}
\item A privacy module that integrates into the GPSD software (runs on every GPS enabled device)
\item A granular privacy interface and control to manage location privacy settings (e.g., location coarseness and release frequency)
\item A performant privacy module with minimal overhead
\end{enumerate}

We first describe the architecture and flow of GPSD, we then describe our privatization algorithms, then we describe our integration with GPSD, and finally we evaluate our scheme.

\section{Related Work}

GPSD is a daemon that network enables the GPS sensor on the majority of mobile embedded systems including Android, iOS, Windows Mobile, UAVs, and driverless cars~\cite{gpsd}. On smartphones the network access is limited to localhost applications only (as opposed to remote applications). GPSD enables unfettered access to location data and does not enable or provide any privacy guarantees. \projecttitle provides a privacy module that provides uniform private access across all platforms.

Mobile device permission systems has received attention in the past. Human interaction studies which seek to enhance reader comprehension have been proposed and evaluated~\cite{DBLP:conf/soups/FeltHEHCW12,DBLP:conf/ccs/FeltEW12}. Such systems lack strong and enforacable privacy guarantees. Static analysis tools have been proposed~\cite{DBLP:conf/ccs/YangYZGNW13}. Though such systems serve only to notify the data owner of privacy breaches and are unable to enforce any privacy runtime guarantees. However, these solutions modify the underlying OS thus making them specific to a single OS or device~\cite{DBLP:conf/trust/ZhouZJF11,xprivacy}. Furthermore, these solutions are unable to balance the privacy and utility tradeoff, ultimately resulting a binary approach to privacy.

To guarantee data owner privacy upon the release of data, various mechanisms have been proposed \cite{DBLP:conf/icde/MachanavajjhalaGKV06,Sweene02,LiLV07,DBLP:conf/icalp/Dwork06,DBLP:conf/tcc/DworkMNS06}. Differential privacy has emerged as the strongest of these privacy mechanisms~\cite{DBLP:conf/icalp/Dwork06,DBLP:conf/tcc/DworkMNS06}. The core idea of differential privacy is to provide strong bounds and guarantees on the privacy leakage when multiple aggregate analytics are run despite the presence or absence of a single data owner from the dataset. This privacy mechanism is provided by adding differentially private noise to the aggregrate answer. As opposed to the originally proposed differentially private mechanism which first collects data in a centralize database and then privatizes the release of the data, \projecttitle immediately privatizes the data at the data source (sensor) in real-time.
\section{Goals and Problem Statement}

We now describe the system goals, performance goals, threat model, and privacy goals of \projecttitle. 

\subsection{System Goals}

There should be well defined and enforced constraints regarding third party application's (apps) access to location data. The data owner should be able to specify the constraints such as how accurate location information should be disclosed and how frequent the location data should be disclosed.

Apps only have access to the privatized data and are unable to directly access GPSD daemon and data. All location data released must be approved by the data owner.

The system should support applications that need real-time access to location data. The privacy policy defines how frequently the application is allowed to receive updates (express in epochs), how accurate the location data may be, and geographical regions as to where the application is allowed to receive location data from.

We use a social network messaging application as an example. The application may want to know which city an individual is in, though pinpoint location information within meter accuracy is not required. The data owner is allowed to define both the radius (e.g., city) that is allowed to be returned as well as the frequency (e.g., say at most every hour).

Ultimately the data owner has final say over how location data and the tradeoff between privacy and utility. The utility has benefits for third party analysts interesting in learning aggregate behavior.

\subsection{Performance Goals}

The system should scale gracefully as the number of applications connecting to the GPSD daemon increases. Location data  will be released within the defined epochs.

\subsection{Threat Model}

Mobile devices (e.g., smartphones, tablets, wearables) are under the data owner's control. Kernel and underlying OS is vetted and verified (signatures and trusted sources). Focus is not on low level system threats. We assume that the operating system itself is not malicious and provides a mechanism to provide a privacy policy settings manager accessible to the data owner. Secure micro kernels such as seL4 address these issues and are out of scope for this paper. Applications do not have a system exploit (e.g., rootkit) to circumvent the system.

Applications may try to request data more frequently than the defined epoch. \projecttitle will deny such aggressive requests and ensure that data is only released within the defined epoch.

Applications may act as sybils and send false application IDs in order to confuse the GPSD daemon. \projecttitle will treat sybil applications accordingly using data owner defined defaults. Thus, sybil applications may either be receive location data  using default privacy configurations or not at all.

\subsection{Privacy Goals}

Data owners should be able to limit how frequently an application access location data. Data owners should also be able to define fine-grained access to location data. Applications for which the data owner feels the application does not meter level accuracy, the data owner should be allowed to define a radius from which the location value can be returned from. Additionally, for scenarios where fine-grained location is required, the data owner can define a grid system from which potential locations can be returned from.

GPS sensor data is only accessible via GPSD.

\begin{figure}[t!]
\includegraphics[width=1\columnwidth]{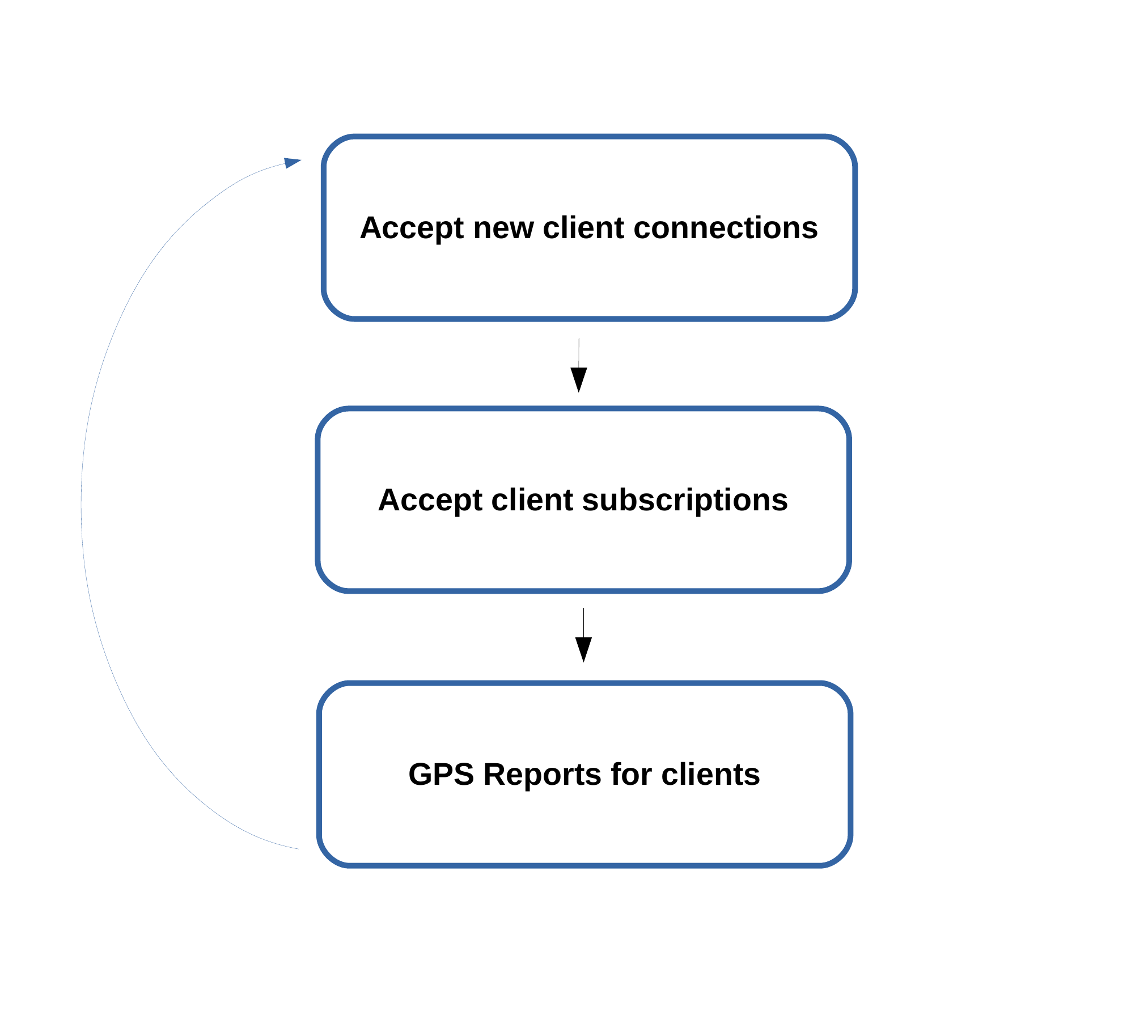}
\caption{GPSD event loop. Privatization occurs when reporting GPS data to the client.}
\label{fig:eventloop}
\end{figure}

\begin{figure}[t]
\includegraphics[width=1\columnwidth]{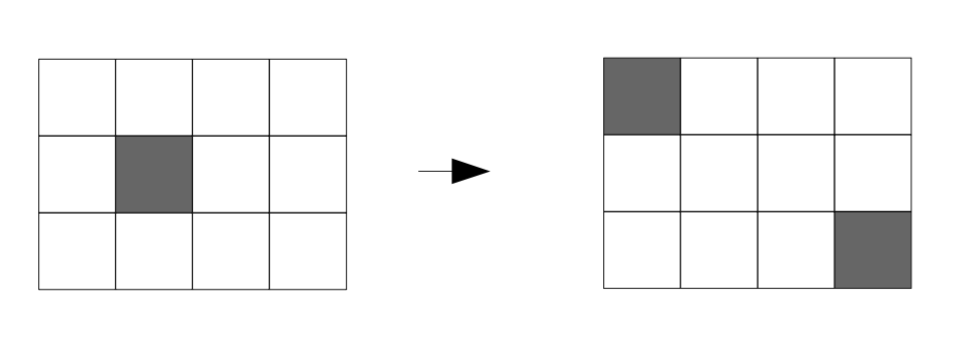}
\caption{In the grid privatization a single location may randomize to one or many locations. In the example above two locations are returned. However, in the aggregate the analyst is able to estimate the underlying population value without violating individual privacy.}
\label{fig:gridlayout}
\end{figure}

\section{Architecture}

Figure~\ref{fig:eventloop} depicts the main components GPSD event loop: accepting new client connections, accepting new client subscriptions, and GPS reporting to all subscribed clients. Each client connecting passes in an application identifier which is mapped to a privacy configuration managed by the system. The privacy configuration contains the epoch (how often data is released in milliseconds), differential private $\epsilon$, privatization radius in meters, and the randomized response coin flips. Clients are allowed to pass in a recommended set of privacy parameters, though these are checked against user settings and are not allowed to exceed the privacy threshold defined by the user. In such cases user settings are adhered to.

\subsection{Privatization}

\projecttitle currently supports two modes of privatization: radius privacy and grid privacy via differential privacy. 

The first mechanism is via radius privacy whereby the data owner can specify a radius cover wherein a random point within the defined range is chosen. This approach favors strong privacy at the expense of utility. That is a larger radius grants more privacy though limits the location accuracy. 

The second privacy mechanism represents the location space as a grid. The grid can be sized according to the data owner's specification. The current location is placed within the grid. Then leveraging the randomized response method one or many grid locations are returned as seen in Figure~\ref{fig:gridlayout}. 

Randomized response~\cite{warner1965randomized} was originally created by social scientists as a mechanism to perform a population study over sensitive attributes (such as drug use or certain ethical behaviors). Randomized response allows data owners to locally randomize their truthful answer to analyts' sensitive queries and respond only with the privatized (locally randomized) answer. We utilize randomized response as our privacy mechanism as randomized response satisfies the differential privacy guarantee for individual data owners, it provides the optimal sample complexity for local differential privacy mechanisms~\cite{DBLP:conf/nips/DuchiWJ13}, and it easily suitable for the location grid type answers we provide.

\subsubsection{Mechanism Description} We will now describe how each data owner privatizes their response utilizing the randomized response mechanism. Suppose each data owner has two independently biased coins. Let the first coin flip heads with probability $p$, and the second coin flip heads with probability $q$.  Without loss of generality, in this paper, heads is represented as ``yes'' (i.e., 1), and tails is represented as ``no'' (i.e., 0).

Each data owner flips the first coin. If it comes up heads, the data owner responds truthfully; otherwise, the data owner flips the second coin and reports the result of this second coin flip.  

Suppose there are $N$ data owners participating in the population study. Let $\hat{Y}$ represent the total aggregate of ``yes`` randomized answers. The estimated population with the sensitive attribute $Y_A$ can be computed as:
\begin{equation}
\label{eqn:yo}
Y_A = \frac{\hat{Y} - (1 - p) \times q \times N}{p}
\end{equation}

The intuition behind randomized response is that it provides ``plausible deniability'', i.e., any truthful answer can produce a response either ``yes'' or ``no'', and data owners retain strong deniability for any answers they respond. If the first coin always comes up heads, there is high utility yet no privacy. Conversely, if the first coin is always tails, there is low utility though strong privacy. It has been shown that by carefully controlling the bias of the two coin flips, one can strike a balance between utility and privacy ( Table 4 in \cite{fox1986randomized} and Table I in \cite{2016arXiv160404810J}).

\subsubsection{Multiple Sensitive Attributes}

While randomized response is an intuitive privacy mechanism for a single location, naturally the question becomes how does one deal with multiple locations, i.e., a grid representation?  A host of "polychotomous" mechanisms have been studied and surveyed in the literature \cite{fox1986randomized} using multiple randomizing mechanisms or maximum likelihood estimators \cite{doi:10.1080/01621459.1981.10477741}. However, it turns out that simply repeating an application of \cite{fox1986randomized} for each grid location turns out to be an ``optimal'' \cite{doi:10.1080/01621459.1981.10477741} approach.

Thus, \projecttitle repeats the randomized response mechanism for each grid location. For example, if a traffic analyst wishes to understand the traffic flow of a few key locations, the traffic analyst issues a query that is a Boolean bit-vector asking each data owner to indicate the location they are at. Then, each data owner performs randomized response for each location and replies with a Boolean bit-vector. The traffic analyst then aggregates and sums the bit-vectors to calculate the number of vehicles at each location.

\section{Evaluation}

\begin{table}[]
\centering
\begin{tabular}{ll|llll}
			 & & \multicolumn{3}{c}{Epoch (seconds)}  &  \\
			 &                      & 5 & 10    & 15 &  \\\hline
\multirow{2}{*}{\# Clients} 	& 25  & 7  & 12  & 16   &  \\
		 					& 64  &  6 & 11  &  14  &  \\
						 	&                      &   &       &    & 
\end{tabular}
\caption{Scaling performance of clients receiving a response in specified epoch. Values are averaged across ten iterations.}
\label{fig:scaling}
\end{table}

To evaluate the overhead of the addition of the privacy module to GPSD, we run 25 and 64 clients connecting to GPSD with varying epochs of 5,10,15 seconds as seen in Table~\ref{fig:scaling}. The evaluation was run on a laptop running Archlinux release 2016.06.01 kernel 4.5.4 with two i5 physical cores (four logical) and 12gb ram. GPSD by default has a limit of 64 clients so we stay within this bound.

The results show that minimal overhead is incurred by the privacy module and that clients are able to reasonably receive location updates within the allotted epoch. Even as more clients connect the performance guarantees do not degrade.

\section{Conclusion}

In this paper we present to our knowledge the first software privacy module for GPSD which is a GPS daemon running on the majority of mobile embedded systems today. Data owners are able to express privacy consent and control by enforcing privacy at the lower level of the OS with minimal runtime overhead.

For future work we plan integration with Android and iOS. This will allow us to evaluate the impact and design on location based services.

%
\bibliographystyle{abbrv}
\bibliography{gpsd,permissions,privacy}  
%
%
\end{document}